# Un esercizio di storia della scienza: misurare i periodi dei satelliti galileiani di Giove con il Sidereus Nuncius

Davide Neri
Liceo A.B. Sabin, Bologna (retired)

**Abstract**
After Galileo's publication of the *Sidereus Nuncius* in 1610, Giovanni Battista Agucchi obtained in 1611 an estimate of the orbital periods of the Galilean satellites of Jupiter using the drawings published in the book. The article shows how to repeat these measurements in a teaching context for high school students.

**Introduzione**
La didattica della storia della scienza è un problema complesso perché richiede la capacità di leggere gli stessi fenomeni e gli stessi problemi con due "occhiali" diversi: quelli di oggi e quelli di un'altra epoca. A questa difficoltà, che accomuna docenti e studenti, si aggiunge, per chi insegna, il problema di presentare in una forma rigorosa, ma al tempo stesso accessibile per gli studenti, una evoluzione delle conoscenze scientifiche che quasi mai è avvenuta in modo lineare, bensì attraverso una rete di questioni e riflessioni che solo in modo superficiale si possono ridurre a qualcosa di semplice. E, spesso, questa semplificazione si traduce in una falsificazione del processo storico.
A titolo di esempio, si dovrebbe evitare che la storia, raccontata col senno di poi, portasse a credere che la Teoria della relatività ristretta di Einstein è figlia dell'esperimento di Michelson-Morley (cosa che lo stesso Einstein ha ripetutamente negato), o che la Teoria quantistica abbia avuto origine dal tentativo di rimediare alla catastrofe ultravioletta (scoperta solo nel 1905, cinque anni dopo la formulazione della teoria di Planck del corpo nero).
Vi sono però casi in cui è possibile calarsi senza eccessive difficoltà in un particolare periodo storico per comprendere le domande a cui gli scienziati cercavano di rispondere. Uno di questi casi riguarda le prime osservazioni telescopiche dei satelliti Giove e la determinazione dei loro periodi; un problema a cui non solo Galileo, ma anche Keplero e altri si dedicarono tra il 1610 e il 1612.
Come è noto, Galileo scoprì i 4 satelliti medicei di Giove (ora denominati Io, Europa, Ganimede e Callisto, dal più interno al più esterno) nei primi giorni del 1610, grazie all'uso di un cannocchiale da lui perfezionato e applicato all'osservazione astronomica, e diede notizia delle sue prime scoperte con la pubblicazione del *Sidereus Nuncius*, avvenuta il 13 marzo 1610. L'impatto che questo libretto ebbe sulla cultura del suo tempo è ben descritto nel volume *Il telescopio di Galileo. Una storia europea*, di M. Bucciantini, M. Camerota, F. Giudice, a cui rinvio per approfondimenti.
Sin dall'inizio, la determinazione dei periodi dei satelliti venne descritta come estremamente complessa. Nel *Sidereus Nuncius* Galileo si limita a dire che il satellite più esterno, attualmente noto come Callisto, compiva una rivoluzione completa attorno a Giove in circa mezzo mese.
Anche Keplero, che poté disporre di un telescopio solo nel settembre dello stesso anno, giudicò in un primo tempo quasi insolubile il problema dei periodi, salvo poi dare valutazioni molto approssimative (14 giorni e 8 giorni, rispettivamente) per i due più esterni e limitandosi a dire che i due più interni hanno periodo più breve, nella prefazione del *Dioptricae*, pubblicato nel 1611:

> tardissima spacio dierum quatuordecim nostratium ut Galilaeus prodidit; proxima ab illa, sed maxime omnium conspicua, spacio dierum octo, ut ego superiori Aprili & Maio deprehendi; reliquae duae multo adhuc breviori temporis curriculo. (in: J. Kepler, *Opera omnia*, 1859, vol. II, p.526)

Tuttavia, nell'*Avvertimento* che introduce al volume terzo, parte seconda, dell'Edizione Nazionale delle *Opere di Galileo* (1890-1904), i curatori segnalano una interessante circostanza. Con il solo aiuto delle figure riportate nel *Sidereus Nuncius*, un corripondente di Galileo, Giovanni Battista Agucchi[1], nel 1611 era riuscito a determinare con buona approssimazione tali periodi, al punto che Galileo stesso, ancora alla ricerca di una loro valutazione esatta, temette di perdere la priorità sulla misura dei periodi e si affrettò a pubblicare i suoi risultati all'inizio del 1612.

Agucchi comunicò a Galileo i risultati ottenuti in una lettera del 14 ottobre 1611:

> … da quanto ella mi ha significato […] sono stato invitato a considerare attentamente i luoghi osservati di essi [i satelliti], che si trovano nel suo Nuntio Sidereo e m'è paruto che se ne possano trarre da vicino […] i periodi delle Stelle.
> Perciò, havendole riconosciute e distinte tutte quante ad una ad una, ho raccolto, che la prima […] fa suo giro in ispatio d' un giorno et hore diciotto et un terzo, o poco poco più parendomi che in giorni sette et hora una e mezza ella il compia quattro volte, con picciola differenza dal più al meno. E la seconda mi mostra che 'l faccia in giorni tre et hore quindici, due volte girandolo in giorni sette et un quarto o in poco manco. Della terza poi […] ho stimato che sia il periodo giorni sette et hore quattro in circa, sì che ella vi spenda quasi il doppio del tempo che v'impiega la seconda […]. L'ultima finalmente mi sembra che si rivolga intorno all' orbe in giorni sedici et hore venti. (*Le opere di Galileo Galilei*, Ed.Naz., vol. XI, p.219)

Dopo aver raccolto molte altre osservazioni, Galileo pubblicò invece i periodi da lui misurati nella primavera del 1612, presentandoli nelle prime pagine del *Discorso intorno alle cose che stanno in su l'acqua*:

> … l'investigazion de' tempi delle conversioni di ciaschedun de' quattro Pianeti Medicei intorno a Giove, la quale mi succedette l'aprile dell'anno passato 1611, mentre era in Roma; dove finalmente m'accertai, che 'l primo, e più vicino a Giove, passa del suo cerchio gradi 8 e m. 29 in circa per ora, faccendo la 'ntera conversione in giorni naturali 1 e ore 18 e quasi mezza. Il secondo fa nell'orbe suo g. 4, m. 13 prossimamente per ora, e l'intera revoluzione in giorni 3, or. 13 e un terzo incirca. Il terzo passa in un' ora gr. 2, m. 6 in circa del suo cerchio, e lo misura tutto in giorni 7, ore 4 prossimamente. Il quarto, e più lontano degli altri, passa in ciaschedun' ora gr. 0, m. 54 e quasi mezzo, del suo cerchio, e lo finisce tutto in giorni 16, or. 18 prossimamente. (*Le opere di Galileo Galilei*, Ed.Naz., vol. IV, p.63)

C'è da notare, *en passant*, che Galileo fa risalire le sue valutazioni alla primavera del 1611, un anno prima della loro pubblicazione e alcuni mesi prima della lettera di Agucchi.
Questa retrodatazione dalla primavera del 1612 a quella del 1611, dichiarata da Galileo e della quale non ci sono prove certe nei manoscritti, ha fatto sospettare ai curatori dell'Edizione Nazionale che Galileo volesse rivendicare una priorità temporale sulle misure dei periodi pur senza averne un pieno diritto[2].
Ci si può rendere conto della qualità delle diverse stime dei periodi confrontando i valori di Agucchi e quelli di Galileo con quelli attuali.

---

[1] G.B. Agucchi (1570-1632) fu un ecclesiastico e diplomatico pontificio con svariati interessi letterari, scientifici e artistici. Conobbe Galileo a Roma nella primavera del 1611 ed entrò in corrispondenza con lui.

[2] Il problema della priorità è sempre stato importante, ma in quel contesto, di fronte a scoperte che promettevano di sconvolgere l'intera visione del mondo, diventava ancor più fondamentale, perché dalla priorità derivava l'*autorità* di chi la poteva rivendicare. Infatti Galileo fu molto contrariato quando i Gesuiti, dopo aver ammesso la validità delle sue osservazioni telescopiche, insinuarono che qualcuno di loro le aveva fatte prima di lui (cfr. Bucciantini e altri, *Op. cit.*, pag.228).

| Satellite | Agucchi (autunno 1611) | Galileo (primavera 1611?-1612?) | Valori attuali [3] |
|---|---|---|---|
| Io | 1d 18h 20m | 1d 18h 30m | 1d 18h 27m 34s |
| Europa | 3d 15h | 3d 13h 20m | 3d 13h 13m 42s |
| Ganimede | 7d 4h | 7d 4h | 7d 3h 42m 33s |
| Callisto | 16d 20h | 16d 18h | 16d 16h 32m 11s |

La corrispondenza è notevole, soprattutto se si tiene conto che le misure delle distanze dei satelliti da Giove nei primi due anni furono fatte, come si afferma nel *Discorso*, "a occhio". Solo a partire dal 31 gennaio 1612 Galileo iniziò ad usare un micrometro, cioè un regolo graduato (o un reticolo), montato sul cannocchiale in modo da poter scorrere lungo il tubo e illuminato da una candela. Guardando con un occhio nel tubo e con l'altro il micrometro, si poteva ottenere un'immagine sovrapposta che consentiva di leggere le distanze tra Giove e i satelliti, misurate in semidiametri di Giove, utilizzando quindi una scala che non dipendeva dalla distanza del pianeta dalla Terra.

Il punto che voglio sottolineare, e che è oggetto di questo articolo, è tuttavia questo: se Agucchi ha ottenuto risultati soddisfacenti con i soli dati forniti dalla stampa del *Sidereus Nuncius*, che si possono ancora facilmente reperire nelle edizioni moderne dell'opera o in quella originale, allora è possibile ripetere la sua esperienza in un contesto didattico.

**Problemi di osservazione**
L'osservazione dei satelliti di Giove dalla Terra pone alcuni problemi di "prospettiva" che possono ostacolare la comprensione dei dati. Noi vediamo i satelliti di Giove solo quando si verificano due condizioni: a) sono illuminati dal Sole; b) Giove non li copre alla nostra vista. Pertanto, il satellite in figura 1 può essere invisibile perché è occultato da Giove, che si pone tra il satellite e la Terra, oppure perché è eclissato da Giove, che si pone tra il satellite e il Sole. Ai casi a) e b) si aggiungeva, per i telescopi di Galileo, il caso del transito c) con il satellite che si trova tra Giove e la Terra.

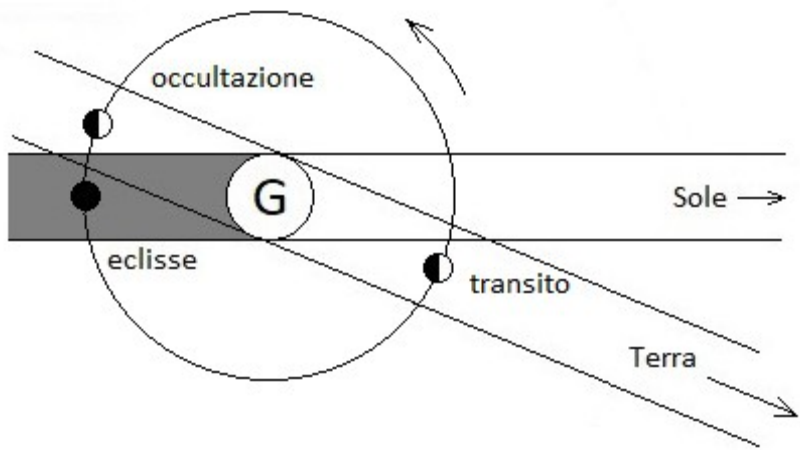

Figura 1

Per la rilevazione delle posizioni, i casi in cui i satelliti, pur essendo potenzialmente visibili, di fatto non lo erano perché eclissati da Giove, pongono problemi aggiuntivi a quelli, più ovvi, dovuti alla loro occultazione. Può accadere che un satellite non si veda nonostante il cammino ottico satellite-Terra sia privo di ostacoli. A quanto pare Galileo cominciò a rendersi conto del problema delle eclissi solo nel marzo del 1612[4], mentre prima attribuiva la non visibilità di certi satelliti, quando essa era invece prevista, all'eccessivo fulgore del pianeta cui erano prossimi. Il lavoro di rifinitura dei dati continuò comunque per vari anni, fino alla fine del 1619[5].

**Come presentare le osservazioni**
Quello dei satelliti attorno a Giove è un moto (quasi) circolare uniforme che, a causa del nostro punto di osservazione, ci appare come un moto armonico semplice unidimensionale. Se nella rappresentazione di tale moto introduciamo il tempo, otteniamo una sinusoide che esprime la variazione delle posizioni del corpo studiato in funzione di *t* (vedi figura 2).
Se si scorre il *Sidereus Nuncius*, nella parte dedicata ai satelliti di Giove si incontra una lunga serie di osservazioni, rappresentate con caratteri tipografici, che vanno dal 7 gennaio al 2 marzo 1610. A titolo di esempio, in figura 3 mostro le prime tre, del 7, 8 e 10 gennaio, ricavate dalla prima edizione, reperibile su Internet Archive. Il cerchio è Giove, gli asterischi sono i satelliti visibili.

---
3 Dati reperibili su Wikipedia (versione italiana) *Satelliti naturali di Giove*, e altri siti, con discordanze trascurabili.
4 *Le opere di Galileo*, Ed. Naz.,1890-1904, vol. III, parte seconda, pag.422.
5 *Le opere di Galileo*, Ed. Naz.,1890-1904, vol. III, parte seconda, pag.423.

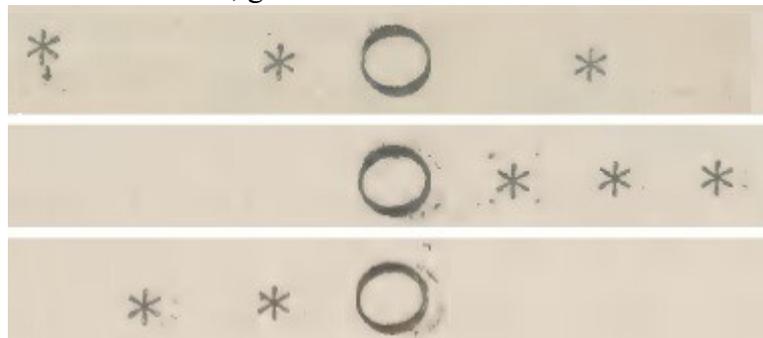

Figura 3

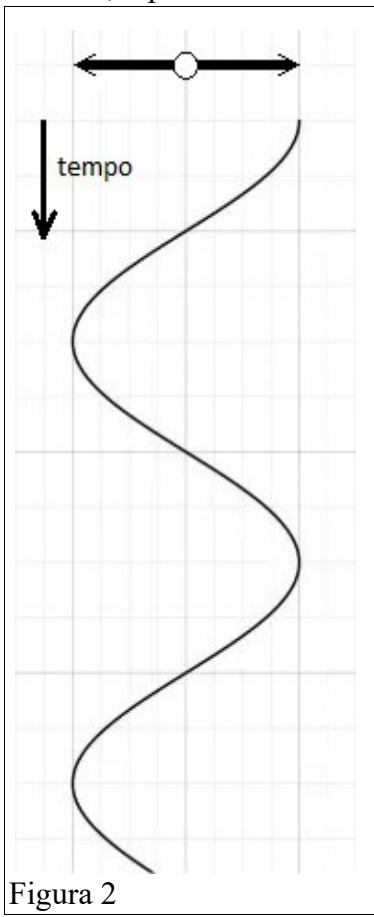

Figura 2

Per ottenere una rappresentazione simile a quella della sinusoide di fig. 2, viene dunque spontaneo riprodurre una dopo l'altra le osservazioni di Galileo su un foglio di carta a quadretti, in cui ogni riga corrisponde a un giorno, avendo ovviamente cura di lasciare bianche le righe dei giorni in cui non ci furono osservazioni. L'uso di questa procedura può sembrare arcaico, ma si avvicina molto a quella che può essere stata usata da Agucchi per fare le sue valutazioni.

Le osservazioni di Galileo sono sempre accompagnate dalla data e, spesso ma non sempre, dall'ora di notte in cui sono state compiute. Si può presumere, quando manca un'indicazione, che siano state compiute nelle prime ore dopo il tramonto. C'è da notare che nel *Sidereus Nuncius* Galileo conta le ore a partire dal tramonto. Ciò significa che l'ora variava coi giorni dell'anno: a Padova, luogo in cui Galileo si trovava, all'inizio di gennaio il Sole tramonta intorno alle nostre 16.45, all'inizio di febbraio alle 17.15, all'inizio di marzo alle 18 circa. Pertanto, l'indicazione "un'ora dopo il tramonto" poteva significare 17.45, oppure 18.15, o 19, e così via, a seconda del giorno. In alcune notti Galileo fece più osservazioni, indicando sempre l'ora di queste ultime. Se ci si limita a considerare solo le osservazioni singole e le prime osservazioni dei giorni in cui Galileo ne fece più di una, si può con buona approssimazione considerarle compiute a distanza di 24 ore circa l'una dall'altra.

Seguendo il criterio appena esposto, gli studenti possono copiare su un foglio protocollo a quadretti le posizioni dei satelliti con la loro distanza dal centro di Giove, assegnando una riga a ciascun giorno compreso tra il 7 gennaio e il 2 marzo. Nei giorni senza osservazioni la riga deve rimanere vuota. Le prime righe dello schema generale, corrispondenti alla figura 3 ricavata dal *Sidereus Nuncius*, assumono quindi la forma mostrata nella figura 4.

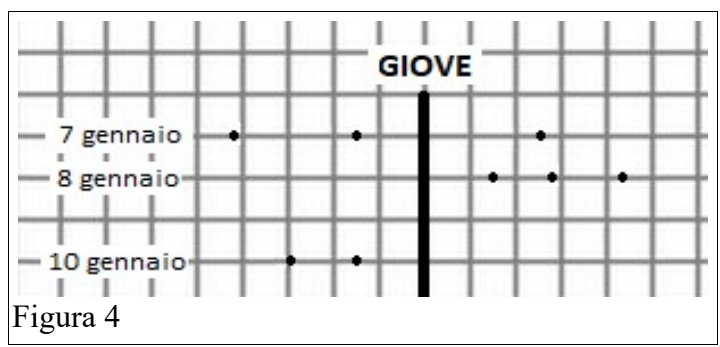

Figura 4

Una volta completato il diagramma appare chiaro quanto fossero prese "a occhio" le misure delle prime distanze. Per esempio: Callisto viene a trovarsi alla sua massima distanza alla sinistra (a oriente) di Giove nei giorni 25-26 gennaio, 10-11 febbraio e 26-27 febbraio, ma in queste due ultime notti la sua distanza dal pianeta, stimata da Galileo, è di circa 13 semidiametri di Giove, mentre nelle coppie di notti precedenti era di circa 16 semidiametri. Si tratta di una differenza notevole, ma non sufficiente a impedire la valutazione dei periodi da parte di Agucchi.

**L'analisi dei dati**

Completata questa trascrizione delle posizioni sul foglio protocollo a quadretti, si ricopia lo schema su un foglio di carta millimetrata trasparente sovrapposto a quello protocollo e si inizia la ricerca delle sinusoidi procedendo dal satellite più esterno (Callisto) fino al più interno. Una volta individuate le posizioni riferibili a Callisto, si cerca di tracciare la sinusoide. Questa prima fase è abbastanza semplice, e aiuta a capire perché sia Galileo, sia Keplero, stimarono rapidamente il periodo di Callisto in circa 15 giorni.

Una volta individuate le probabili posizioni di Callisto, si ricopiano su un altro foglio di carta millimetrata trasparente solo le tracce riconducibili agli altri satelliti. Da queste, si cerca di individuare quelle riferibili a Ganimede, il terzo satellite in ordine di distanza, collegandole con una nuova sinusoide. Questo procedimento si ripete, con nuovi fogli di carta millimetrata, per le tracce riferibili a Europa e Io, determinando prima la sinusoide del secondo satellite, e infine quella del primo, il più vicino a Giove dei quattro.

Come al solito, dovendo determinare un periodo, conviene misurare il tempo che intercorre tra più oscillazioni complete e dividerlo, riducendo così l'incertezza relativa. A quanto pare Agucchi e Galileo hanno proceduto allo stesso modo, prendendo un numero intero (o semintero) di giorni e dividendolo per un numero intero di periodi. Faccio alcuni esempi:

- il periodo di Europa misurato da Galileo corrisponde esattamente a 9 periodi in 32 giorni; infatti T = (32/9)d = $3,\bar{5}$d = 3d 13h 20m, quello misurato da Agucchi corrisponde a 8 periodi in 29 giorni; infatti T = (29/8)d = 3⅝d ≅ 3d 15h;
- il periodo di Ganimede misurato da Agucchi e Galileo corrisponde a 6 periodi in 43 giorni; infatti T= (43/6)d = 7⅙d = 7d 4h.

La valutazione da me fatta utilizzando numeri interi di giorni ha dato invece questi risultati:

- CALLISTO compie 3 rivoluzioni complete in circa 50 giorni (dal 9 gennaio al 28 febbraio), pertanto $T_{CALLISTO}$ ≅ 16⅔ d ≅ 16d 16h;
- GANIMEDE compie 7 rivoluzioni in circa 50 giorni (dall'11 gennaio al 2 marzo), pertanto $T_{GANIMEDE}$ ≅ 7⅐ d ≅ 7d 3h 25m;
- EUROPA compie 8 rivoluzioni in circa 29 giorni (dal 31 gennaio all'1 marzo), pertanto $T_{EUROPA}$ ≅ 3⅝d ≅ 3d 15h (lo stesso di Agucchi)
- Io compie 8 rivoluzioni in circa 14 giorni (dal 30 gennaio al 13 febbraio), pertanto $T_{IO}$ ≅ 1¾ ≅ 1d 18h.

**Conclusione**

La bontà della valutazione può essere controllata estrapolando i risultati ottenuti ai giorni esterni all'intervallo utilizzato: si dovrebbe trovare una corrispondenza accettabile tra la sinusoide calcolata e le posizioni osservate. Ovviamente un errore nella determinazione del periodo porta a sfasamenti tra previsione e osservazione tanto più grandi quanto più ci si allontana dall'intervallo utilizzato per il calcolo. Nel caso dei valori da me ottenuti, questa estrapolazione è possibile solo per i due satelliti più interni, in quanto per quelli più esterni il tempo totale utilizzato per la stima dei periodi (50 giorni) abbraccia quasi tutte le osservazioni presentate nel *Sidereus Nuncius*. D'altra parte, sono proprio le stime dei periodi di Europa e Io a presentare le maggiori difficoltà. Nei limiti imposti dalle date di osservazione del *Sidereus Nuncius*, i periodi qui calcolati mostrano un buon accordo con le osservazioni. La tabella seguente riassume i risultati ottenuti e li confronta con quelli di Agucchi e Galileo e con quelli attualmente accreditati:

| Satellite | Agucchi | Galileo | Valori attuali (Wikipedia) | Valori qui ottenuti |
|---|---|---|---|---|
| Io | 1d 18h 20m | 1d 18h 30m | 1d 18h 27m 34s | 1d 18h |
| Europa | 3d 15h | 3d 13h 20m | 3d 13h 13m 42s | 3d 15h |
| Ganimede | 7d 4h | 7d 4h | 7d 3h 42m 33s | 7d 3h 25m |
| Callisto | 16d 20h | 16d 18h | 16d 16h 32m 11s | 16d 16h |

Al termine dell'attività gli studenti potrebbero ricostruire una "mappa" delle osservazioni del *Sidereus Nuncius* che assegna a molti punti segnati sullo schema, se non a tutti, il nome del satellite corrispondente, confrontando tra loro i risultati ottenuti.

**Bibliografia**